# Morphological Stability of Solid-Liquid Interfaces Under Additive Manufacturing Conditions

Damien Tourret[1], Jonah Klemm-Toole[2], Adriana Eres Castellanos[2], Brian Rodgers[2], Gus Becker[2], Alec Saville[2], Ben Ellyson[2], Chloe Johnson[2], Brian Milligan[2], John Copley[2], Ruben Ochoa[2], Andrew Polonsky[3], Kira Pusch[3], Michael P. Haines[4], Kamel Fezzaa[5], Tao Sun[5], Kester Clarke[2], Suresh Babu[4], Tresa Pollock[3], Alain Karma[6], Amy Clarke[2*]

[1] IMDEA Materials, Madrid, Spain
[2] Colorado School of Mines, Golden, CO, USA
[3] University of California Santa Barbara, Santa Barbara, CA, USA
[4] University of Tennessee Knoxville, Knoxville, TN, USA
[5] Advanced Photon Source, Argonne National Laboratory, Lemont, IL, USA
[6] Northeastern University, Boston, MA, USA
* Email: amyclarke@mines.edu

## Abstract

Understanding rapid solidification behavior at velocities relevant to additive manufacturing (AM) is critical to controlling microstructure selection. Although *in-situ* visualization of solidification dynamics is now possible, systematic studies under AM conditions with microstructural outcomes compared to solidification theory remain lacking. Here we measure solid-liquid interface velocities of Ni-Mo-Al alloy single crystals under AM conditions with synchrotron X-ray imaging, characterize the microstructures, and show discrepancies with classical theories regarding the onset velocity for absolute stability of a planar solid-liquid interface. Experimental observations reveal cellular/dendritic microstructures can persist at velocities larger than the expected absolute stability limit, where banded structure formation should theoretically appear. We show that theory and experimental observations can be reconciled by properly accounting for the effect of solute trapping and kinetic undercooling on the velocity-dependent solidus and liquidus temperatures of the alloy. Further theoretical developments and accurate assessments of key thermophysical parameters – like liquid diffusivities, solid-liquid interface excess free energies, and kinetic coefficients – remain needed to quantitatively investigate such discrepancies and pave the way for the prediction and control of microstructure selection under rapid solidification conditions.

## 1. Introduction

The ability to accurately understand, predict, and control solidification behavior in metals and alloys is crucial to achieve advanced manufacturing. Historically, theoretical and experimental analyses of solidification have focused on relatively slow to moderate solid-liquid interface velocities relevant to directional solidification, experienced for instance during casting processes. The fundamental understanding of microstructure selection at high velocities was primarily built on welding or quenching experiments and subsequent microstructural analyses of solidified samples. Seminal studies have identified crucial phenomena in high-velocity pattern formation, such as banding instabilities [1-4]. More recently, the emergence of additive manufacturing (AM) – relying on technologies that often involve relatively high solidification rates – has renewed the general interest in rapid solidification. In the meantime, the use of techniques for *in-situ* imaging of metals processing has also grown substantially [5-9]. These advances have paved the way for a deeper look and updated analyses on the classical assumptions and models used for the rapid solidification of metals and alloys, based on sound, well-controlled, and monitored experiments. Here we analyze measurements from *in-situ* synchrotron X-ray imaging of simulated laser powder bed fusion (PBF-LB), involving laser melting and rapid solidification of model ternary Ni-Mo-Al alloy single crystals and discuss them in light of existing solidification theories.

For a given alloy composition, the expected solidification pattern development is primarily determined by the solid-liquid interface velocity (V) and temperature gradient (G) [10,11]. For a given temperature gradient, at low velocities, a planar interface is stable. Solute atoms are rejected in the liquid from the solid-liquid interface due to the difference in solubility between the solid and liquid (partitioning). A solute boundary layer builds up ahead of the interface, with a decreasing exponential profile, and accordingly with solute conservation, the solid-liquid interface stabilizes within the temperature gradient at the solidus temperature. With increasing solidification velocity, the solute profile in the liquid is squeezed toward the solid-liquid interface. Above a given velocity, this results in a region ahead of the interface that is undercooled below its local liquidus temperature (i.e., relative to its local concentration). This velocity, $V_c$, is termed the constitutional undercooling limit. A precise calculation of this threshold was performed by Mullins and Sekerka (MS) based on a perturbation analysis [12]. The simple constitutional undercooling argument proposed earlier by Tiller et al. [13] provides a close approximation to the more complete MS analysis. Above $V_c$ the interface breaks down into a cellular pattern, which turns into a dendritic pattern as the velocity increases further.

One way to distinguish cells from dendrites is that cells grow parallel to the primary heat flow direction, whereas dendrites tend to grow along specific crystallographic directions, such as <100> directions close to the heat flow direction in most cubic crystals [10,14,15]. Cells and dendrites grow with their tips located below the liquidus temperature of the alloy. This undercooling is comprised of solute, curvature, and kinetic components [10,11]. With increasing solidification velocity, dendritic arrays become increasingly finer, until reaching a much higher velocity, typically orders of magnitudes faster than $V_c$, at which the interface pattern changes back to a cellular and then planar interface. This critical velocity, $V_{abs}$, is usually called the absolute stability limit [12].

The restabilization of a planar interface is not the only manifestation of rapid solidification. In fact, it is accompanied by two major phenomena. First, as the interface velocity V becomes of the same order of magnitude as the diffusion velocity of atoms through the interface, $V_D$, thermodynamic equilibrium at the interface cannot be maintained, and more solute atoms are incorporated into the solid than dictated by the equilibrium phase diagram. This phenomenon, referred to as solute trapping [16-18], becomes crucial across a velocity range with $V \geq V_D$ and leads to the solidus and liquidus lines of the phase diagram ultimately merging at the so-called $T_0$-line [11]. The second additional manifestation of rapid solidification is kinetic undercooling, which increases proportionally with growth velocity and becomes predominant at high V. Essentially, this leads to a shift of the equilibrium phase diagram toward lower temperatures. Kinetic undercooling, solute trapping, and absolute stability are the three main components of rapid solidification. They are intricately linked to one another and may occur over distinct velocity ranges. In multicomponent alloys, solute trapping can even occur across different velocity ranges for the individual species, such that it is often not trivial to unequivocally identify whether a given regime qualifies as "rapid solidification". For these reasons, whether a given AM process for a given alloy can be rigorously qualified as rapid solidification is often ambiguous.

In the velocity regime between $V_c$ and $V_{abs}$, various models were proposed to estimate the growth temperature and key microstructural features, amongst which the cell/dendrite tip radius plays a prominent role. An elegant – and arguably the most widely used – model was proposed by Kurz, Giovanola, and Trivedi, and is often referred to as the KGT model [19]. The model encapsulates some key ingredients for solidification across a broad velocity range, namely thermal conditions, solute partitioning, interface curvature (capillarity), and solute transport (diffusion). Still, in its simplest form, i.e., for a binary alloy at a given temperature gradient, it consists in finding the roots of a simple second order polynomial. The KGT model yields quite accurate predictions of pattern formation during solidification for low and medium interface velocities. With minor modifications, it has also provided valuable insights into rapid solidification behavior in binary [19] or ternary [20] systems. Yet, while the KGT model remains a theoretical staple commonly used in the context of AM, its validity and limitations for high solidification rates remain to be discussed considering modern, *in-situ* AM-relevant experiments under well controlled and monitored conditions.

Here we present and discuss the results of laser melting and solidification experiments on single crystal nickel-based ternary alloys, in which the solid-liquid interface velocities were measured *in-situ* via synchrotron X-ray imaging. We compare the obtained microstructures with predictions from the classical KGT theory. Resulting discrepancies demonstrate that more advanced theories are needed to appropriately describe the threshold velocities required to reach absolute stability. Before discussing our results, we review the key features and equations of the relevant theories.

## 2. Classical Theories for Dendritic Growth Kinetics

### 2.1. Planar Stability Limits

While the Mullins-Sekerka (MS) analysis [12] provides a more accurate representation of the planar interface instability at $V \approx V_c$, it is most often substituted by the simpler and close approximation based on the diffusion profile ahead of the interface and the liquidus slope of the alloy, i.e., the constitutional undercooling criterion [13]:

$$V_c = \frac{DGk}{m(k-1)c_0} \tag{1}$$

where D is the solute diffusivity in the liquid phase, $k = c_s/c_l$ is the interface solute partition coefficient with $c_s$ and $c_l$, respectively, as the equilibrium solute concentrations on the solid and liquid sides of the interface, m < 0 is the slope of the liquidus line in the alloy phase diagram, and $c_0$ is the alloy nominal concentration. On the other hand, the stability analysis in the high velocity regime leads to the absolute stability limit [12,21]:

$$V_{abs} = \frac{D\Delta T_0}{k\Gamma} = \frac{Dm(k-1)c_0}{k^2\Gamma} \tag{2}$$

where $\Delta T_0$ is the freezing range of the alloy at $c_0$, and $\Gamma$ is the Gibbs-Thomson coefficient of the solid-liquid interface. For ternary or multicomponent alloys, if the additional alloying elements are sufficiently dilute to assume that their interaction is minimal (e.g., no cross-coupling extra-diagonal terms in the diffusivity matrix), one can simply sum up their individual contributions [22] and establish the following expressions (here generalized for N solute elements):

$$V_c = G \Big/ \sum_{i=1}^{N} \left( \frac{m^{(i)}(k^{(i)}-1)c_0^{(i)}}{D^{(i)}k^{(i)}} \right) \tag{3}$$

$$V_{abs} = \frac{1}{\Gamma} \sum_{i=1}^{N} \left( \frac{D^{(i)}m^{(i)}(k^{(i)}-1)c_0^{(i)}}{k^{(i)^2}} \right) \tag{4}$$

where superscripts (i) stand for the different solute species. As can be seen in the equations above, the breakdown of the planar interface at low velocity ($V_c$) strongly depends on the temperature gradient. This breakdown indeed arises from the temperature gradient and solute composition profile in the liquid being of similar length scales. At absolute stability ($V_{abs}$), on the other hand, the solute composition profile is confined to the immediate vicinity of the interface and possibly reduced by solute trapping, such that the temperature gradient effect is marginal. However, Eq. (4) is derived by assuming that the values of $k^{(i)}$ [16,17] and $m^{(i)}$ [23], which take into account the departure of chemical equilibrium at the solid-liquid interface under rapid solidification conditions, have fixed values determined by the steady-state interface velocity V. A rigorous linear stability analysis, which relaxes this assumption by self-consistently taking into account the time-dependent changes of $k^{(i)}$ and $m^{(i)}$ associated with the growth of perturbations of the planar interface, predicts that instability persists for a range of velocities larger than $V_{abs}$ [3], as further discussed below.

## 2.2. Kurz-Giovanola-Trivedi Model

Velocities $V_c$ and $V_{abs}$ predicted by the KGT model are consistent with the expressions above. For $V_c < V < V_{abs}$, the model predicts the cell/dendrite tip radius, R, and its temperature, T, as a function of the growth velocity, V, for a given temperature gradient, G, and a set of alloy parameters – namely $\Gamma$, $k^{(i)}$, $m^{(i)}$, and $c_0^{(i)}$. The original formulation of the model considers a constant diffusivity of solute elements in the liquid, a fixed partition coefficient, and a fixed liquidus slope. A simple extension to a multicomponent alloy using additive contributions of the different species results in the following system of equations [19,20]:

$$\frac{4\pi^2 \Gamma}{R^2} + \frac{2}{R} \sum_i \left[ \frac{Pe^{(i)}m^{(i)}c_0^{(i)}(1-k^{(i)})\xi(Pe^{(i)},k^{(i)})}{1-[(1-k_i)Iv(Pe_i)]} \right] + G = 0 \tag{5}$$

with

$$\xi(Pe, k) = 1 - 2k \Big/ \left( \sqrt{1 + \left(\frac{2\pi}{Pe}\right)^2} - 1 + 2k \right) \tag{6}$$

and the tip temperature given by

$$T = T_M + \sum_i \left[ \frac{m^{(i)}c_0^{(i)}}{1-(1-k^{(i)})Iv(Pe^{(i)})} \right] - \frac{2\Gamma}{R} \tag{7}$$

where $T_M$ is the melting temperature of the pure solvent – sometimes using a fictitious value extrapolated from the liquidus slope of the alloy at its liquidus temperature [20]. The Péclet number for solute species (i) is:

$$Pe^{(i)} = \frac{RV}{2D^{(i)}} \tag{8}$$

and Iv(Pe) is the Ivantsov solution of the steady diffusion field in front of a parabolic tip [24]:

$$\mathrm{Iv(Pe)} \;=\; \mathrm{Pe\,exp(Pe)\,E(Pe)} \tag{9}$$

In this form, the KGT model reduces to a quadratic equation that can be solved analytically. Typical results provide R(V) and T(V), with asymptotes (R → ∞) when V tends to either $V_c$ or $V_{abs}$. The model can be made to account for the temperature-dependence of the diffusion coefficients through the Arrhenius law [19,20]:

$$D^{(i)}(T) \;=\; D_0^{(i)} \exp\left(\frac{-E_D^{(i)}}{R_g T}\right) \tag{10}$$

with $R_g$ the molar gas constant and $E_D^{(i)}$ the activation energy for the diffusion of species (i), and to account for the velocity-dependence of the solute partition coefficient [19,20]:

$$k^{(i)} \;=\; \frac{k_0^{(i)} + V/V_D^{(i)}}{1+ V/V_D^{(i)}} \tag{11}$$

with the subscripts 0 denoting its reference values at V = 0. Equation (11) follows the continuous growth model (CGM) proposed by Aziz and co-workers [16,17]. It features a typical diffusion velocity for species through the interface, $V_D^{(i)}$. This velocity is often estimated as $D^{(i)}/\delta$, with $\delta$ as the width of the solid-liquid interface, on the order of nanometers. Yet, it is more accurately estimated using the expression [25,26]:

$$V_D^{(i)} \;\approx\; 0.207 \left(\frac{\ln\left(1/k_0^{(i)}\right)}{1-k_0^{(i)}}\right) \frac{D^{(i)}}{\delta} \tag{12}$$

Moreover, the model can be further extended to account for the velocity-dependence of the liquidus slope [23,25]:

$$m^{(i)} \;=\; m_0^{(i)} \left[\frac{1 - k^{(i)} + \left(k^{(i)} +(1 - k^{(i)})\alpha\right)\ln\left(k^{(i)}/k_0^{(i)}\right)}{1- k_0^{(i)}}\right] \tag{13}$$

where $0 \leq \alpha \leq 24/35$ is the solute drag coefficient. Notably, the KGT model has also been further extended to multicomponent alloys, including solute cross-coupling terms [27]. The resulting equations are more complex and not as trivial to solve, but they tend to become necessary for accurate calculations as soon as more than one element cannot be considered as dilute anymore.

## 2.3. Effect of Kinetic Undercooling and Banding Instability

One key ingredient missing from the equations above is the presence of kinetic undercooling, which is necessary to capture the growth kinetics of dendrites at high velocities [28,29]. It can be incorporated into the expressions of the alloy solidus and liquidus temperatures:

$$T_S(V) = T_M + \sum_i \left(\frac{m^{(i)} c_0^{(i)}}{k^{(i)}}\right) - \frac{V}{\mu} \tag{14}$$

$$T_L(V) = T_M + \sum_i \left(m^{(i)} c_0^{(i)}\right) - \frac{V}{\mu} \tag{15}$$

where $\mu$ is the interface kinetic coefficient. Using the velocity-dependent expressions of $k^{(i)}$ and $m^{(i)}$, these equations describe the equilibrium liquidus and solidus temperatures for a planar interface. The resulting shape of $T_S(V)$ is usually non-monotonic. Indeed, the change in $k^{(i)}$ (Eq. (11)), predominant at $V \approx V_D$, tends to increase $T_S$ with V, and it competes with the change in $m^{(i)}$ (Eq. (13)) and the kinetic undercooling, predominant at $V > V_D$, which tend to decrease $T_S$ with V. As the growth velocity increases toward the rapid solidification regime, the decrease of driving force (undercooling) with increasing velocity in the initial region with $dT_S/dV > 0$ [1,4], combined with the effect of latent heat rejection at the interface [3], results in a cyclic instability at the origin of banded structures observed post-mortem in a broad range of welding or laser-melting experiments [1,4] and recently observed *in-situ* in rapid solidification experiments [7,9]. In fact, the absolute stability of the planar interface can only be attained when $dT_S/dV < 0$, such that the real absolute stability threshold $V_{abs}$ is actually given by the maximum of the $T_S(V)$ curve [3,4], which occurs further than predicted by the KGT model and Eq. (4).

In what follows, we show that the KGT model is insufficient, even when adjusted to account for the effects of temperature and velocity on D, k, and m, to reproduce experimental observations of simulated AM, while the predictions of absolute stability based on the maximum of $T_S(V)$ appear to be more consistent with experiments.

## 3. Methods

### *3.1. Experiments*

Experimental observations in the velocity regime relevant to absolute stability for metals requires an experimental setup that enables direct measurements of such solidification velocities. *In-situ* observations of solidification were made possible by an AM simulator [30] at beamline 32-ID-B at the Advanced Photon Source at Argonne National Laboratory, combined with high-speed synchrotron x-ray imaging. Melt pools were imaged by synchrotron X-ray radiography at 67200 to 80000 frames per second, with a spatial resolution (pixel size) of 1.93 μm.

Single crystals of a model ternary Ni-based superalloy with a nominal composition of Ni-22.2Mo-2.8Al (wt%) were used for these experiments [31], with either a <100> or <110> crystallographic direction parallel to the laser beam. Both orientations had a <100> direction parallel to the X-ray beam used for imaging (see Fig. 1a). Spot melts were conducted with a 1070 nm fiber laser at powers ranging from 104 to 260 W, a 1 ms pulse duration, and a laser spot size of about 85 μm. Solidification velocities were obtained by measuring the progression of the solid-liquid interface between radiographic snapshots (Fig. 1b) in the direction parallel to a ⟨100⟩ crystal orientation (Fig. 1c). While laser raster experiments could have been more representative of actual AM conditions, single spot melt experiments lead to a reduced extent of convection (although not completely absent), a more homogeneous and symmetrical thermal field, and more accurate measurements of the interface velocity, which are crucial to discuss interface stability at high velocities.

The solidification velocity was extracted from the X-ray radiography videos. We used a spatiotemporal smoothing of the melt pool shape and of its time evolution to reduce the potential error due to the limited contrast and inherent spatiotemporal resolution of the experiment. The procedure relies on two assumptions: (i) during solidification, the melt pool shape can be approximated by an ellipse with horizontal and vertical major axes, and (ii) the time evolution of the melt pool width and depth can each be fitted to a power law. Under these assumptions, we used the following procedure to extract the solidification velocity. First, using a *Python* code (*imageio* library), we enhanced the contrast between solid and liquid by subtracting or dividing (whichever yields the greatest contrast) each image by the previous one and automatically adjusting contrast and brightness. Second, using *ImageJ* (*Fiji*), we manually selected points along the solidification front in snapshots exhibiting a good solid-liquid contrast (from 3 to 10 points per snapshot, between 79 and 882 points per experiment). Thirdly, using *gnuplot*, for each considered snapshot we calculated the melt pool half-width, w(t), and depth, d(t), by fitting the interface points to an elliptical shape of radius:

$$r(\theta, t) = \left( \frac{w(t)^2 d(t)^2}{d(t)^2 \cos^2(\theta) + w(t)^2 \sin^2(\theta)} \right)^{1/2} \quad (16)$$

where $\theta = \tan^{-1}|(y-y_0)/(x-x_0)|$ is the angle from the horizontal axis, and the center of the ellipse $(x_0, y_0)$ is a fixed point in space and time, hence using w(t) and d(t) as fitting parameters. The resulting time evolution of the melt pool half-width and depth were then fitted to power laws $w(t) = w_0 + w_1 t^n$ and $d(t) = d_0 + d_1 t^p$, respectively, where the time origin is taken at the onset of the melt pool shrinking (i.e., of solidification) and fitting parameters are $(w_0, w_1, n)$ or $(d_0, d_1, p)$. These expressions for d(t) and w(t) permit the direct analytical calculation the solidification velocity at any given angle $\theta$ as $V(t) = - d(r(\theta,t))/dt$. More details on the image processing methods, as well as raw and fitted d(t) and w(t) for each experiment are provided in the attached Supplementary Material. Moreover, attached videos show radiographs with and without the resulting final elliptical shape using fitted power laws for d(t) and w(t).

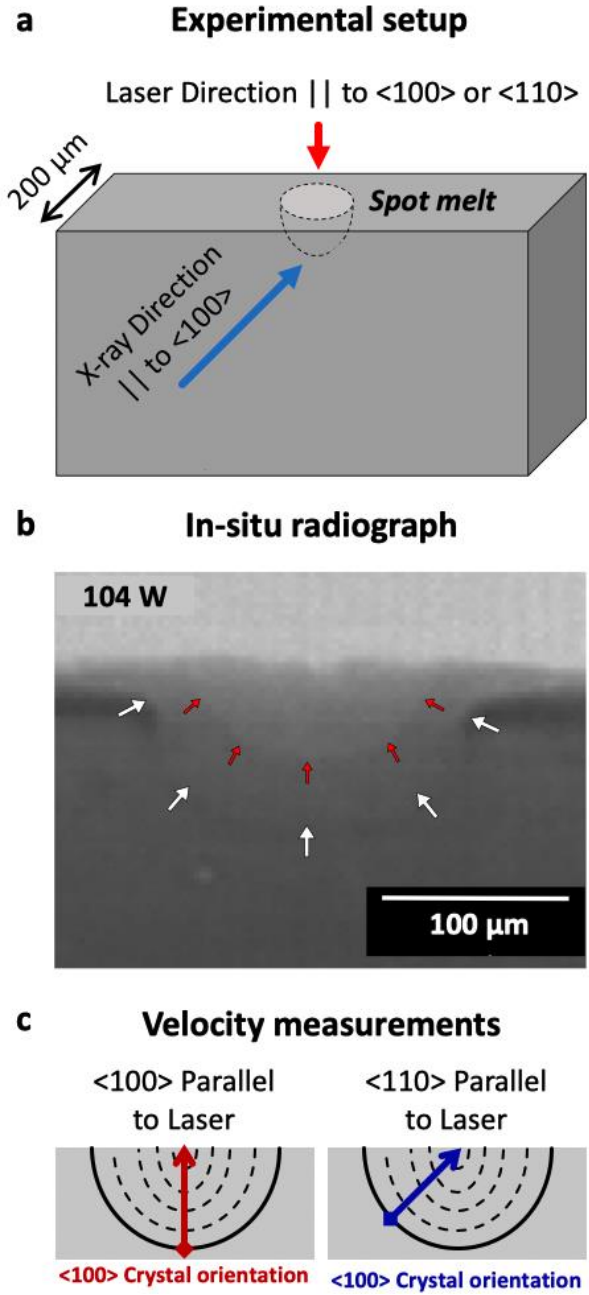

*Fig. 1: (a) Schematic of the experimental set up for in-situ radiography during melting and solidification of single crystal Ni-Mo-Al alloy samples. (b) A synchrotron X-ray radiograph taken during solidification. The white arrows show the starting position of the solid liquid interface, and the red arrows show the position at a later time during solidification. (c) Schematics showing how solid-liquid interface (solidification) velocities were measured along ⟨100⟩ growth directions: solid lines show the fusion line, dashed lines show successive positions of the solidification front, and arrows show the axis over which velocities are measured and plotted in Fig. 3.*

Each solidified sample was also analyzed *ex-situ* using electron microscopy (imaging and diffraction). Each melt pool was imaged from the top-down with scanning electron microscopy (SEM) before any destructive sectioning or polishing. Once top-down imaging was completed, samples were sectioned with a slow-speed saw, hot-mounted, and polished through to the approximate centerline of the melt pool. The depth of material removed was monitored with a ball micrometer. Melt pool centers were imaged through SEM and electron backscatter diffraction (EBSD).

## 3.2. Calculations

The velocity-dependent solidus and liquidus temperatures of a planar interface were calculated from Eqs (14)-(15) using a custom *Python* script, with the local maximum of $T_S(V)$ estimated via libraries *SymPy* and *scipy*. The velocity-dependent dendrite tip temperature, T(V), and radius, R(V), were calculated via the ternary version of the KGT model, using a separate *Python* script, in which the resolution algorithm fixes the velocity and iterates on Eqs (5)-(9) until convergence of the tip temperature between two iterations.

Parameters used in the calculations are listed in Table 1. To the greatest extent feasible, thermophysical parameters of the alloy were assessed from thermodynamic equilibrium calculations using the CalPhaD method. Other parameters were estimated from a thorough review of literature data, both experimental and computational, on Ni-Mo and Ni-Al systems.

Equilibrium concentrations and latent heat of fusion, $L_f$ (used to calculate the Gibbs-Thomson coefficient), were assessed using CalPhaD (ThermoCalc, TCNI8 database). Solute partition coefficients were calculated from CalPhaD-calculated binary Ni-Mo and Ni-Al diagrams. First, each liquidus line was fitted to a polynomial $T_L = T_{Ni} + a\, c^2 + b\, c^3$, which has a null derivative at $c = 0$, where $T_{Ni} = 1728$ K is the melting temperature of pure Nickel, and a and b are fitting parameters. Then, considering the resulting polynomial as the liquidus line, we fitted the CalPhaD-calculated data points along the solidus line, using a constant partition coefficient as single fitting parameter. Both least-square fits were performed along the entire temperature range, exhibiting a stable equilibrium between primary Ni solid phase and liquid. Results of this fitting procedure, as illustrated in Fig. 2a and 2b, show the assumption of constant partition coefficient is accurate, as long as the liquidus line is appropriately captured, at least for the binary alloys.

Once the solute partition coefficients were calculated, we estimated the liquidus slopes by linear fitting of the CalPhaD-calculated liquidus lines within isopleth sections of the ternary phase diagram corresponding to Ni-Mo (at $c^{(Al)}$ = 2.8 wt%Al) and Ni-Al (at $c^{(Mo)}$ = 22.2 wt%Mo), as illustrated in Fig. 2c and 2d. Both fits were performed between solidus ($T_S$) and liquidus ($T_L$) temperatures of the ternary alloy. A (fictitious) melting temperature at $c^{(Al)} = c^{(Mo)} = 0$, $T_M = 1812$ K, was then extrapolated from $T_L$ using the fitted liquidus slopes. The resulting approximations of linearized liquidus and solidus lines appear in Fig. 2c-d. While more accurate linear approximations are possible (e.g., performing a higher-dimension fitting of the entire liquidus and solidus surfaces), Fig. 2c-d show that the resulting linearization provides an excellent approximation of solidus and liquidus lines within the relevant temperature range, i.e., between $T_S$ and $T_L$.

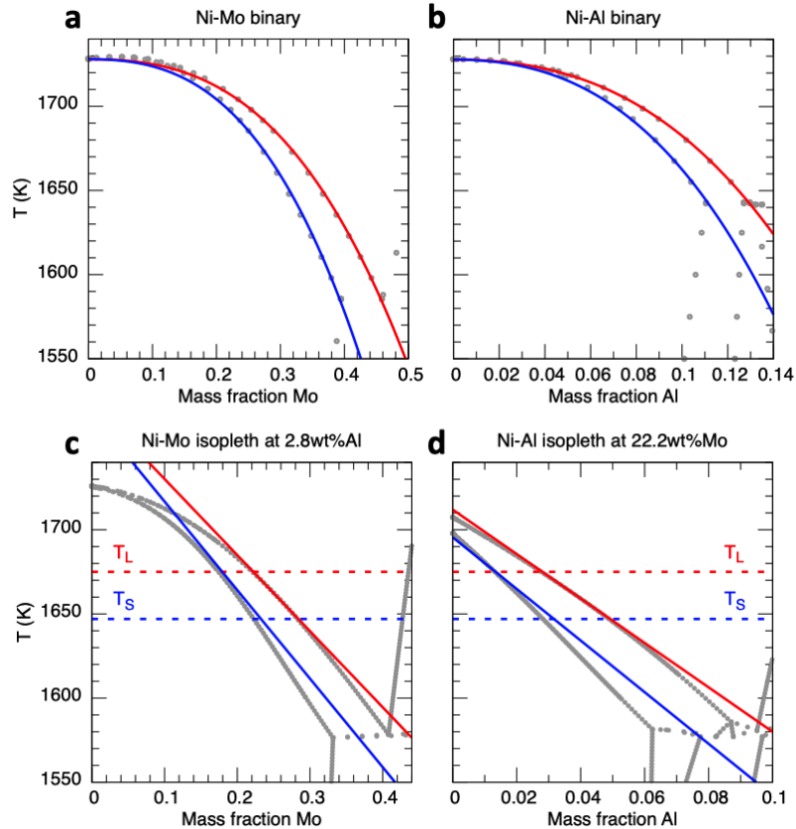

*Fig. 2: (a,b) Polynomial $T_L = T_{Ni} + a\,c^2 + b\,c^3$ approximation of liquidus line (red line) and corresponding solidus line, considering a constant partition coefficient (blue line), both fitted to CalPhaD-calculated phase diagrams (gray dots) for binary Ni-Mo (a) and Ni-Al (b) diagrams. (c,d) Isopleth intersections of the ternary phase diagram (CalPhaD data: gray dots) used for the linear fitting of the liquidus slope between $T_L$ (dashed red line) and $T_S$ (dashed blue line), with resulting linear approximations of liquidus (solid red line) and solidus (solid blue line) using parameters summarized in Table 1.*

Diffusivities for both binary systems were extracted from pulsed ion-beam melting experiments (Ni-Mo [32]) and from specific databases based on compiled experimental data (Ni-Al [33]). The solid-liquid interface energy, $\gamma_0$, used to calculate the Gibbs-Thomson coefficient, $\Gamma = \gamma_0 T_M/L_f$, was assessed for pure Ni from several atomistic studies – independent, yet consistent, with one another [34-36]. The value of the kinetic coefficient of the interface was obtained by molecular dynamics simulations for pure Ni, considering a ⟨100⟩ growth direction [37]. For the only missing parameter, namely the interface width, we considered a realistic order of magnitude with $\delta = 1$ nm.

*Table 1. Alloy parameters considered in the calculations.*

| Parameter | Symbol | Value | Unit |
|---|---|---|---|
| Alloy liquidus temperature | $T_L$ | 1675 | K |
| Equilibrium partition coefficient for Mo | $k_0^{(Mo)}$ | 0.859 | - |
| Equilibrium partition coefficient for Al | $k_0^{(Al)}$ | 0.861 | - |
| Equilibrium liquidus slope for Mo | $m_0^{(Mo)}$ | –4.52 | K/wt%Mo |
| Equilibrium liquidus slope for Al | $m_0^{(Al)}$ | –13.2 | K/wt%Al |
| Solvent melting temperature (fictitious [20]) | $T_M$ | 1812 | K |
| Gibbs-Thomson coefficient of the s/l interface | $\Gamma$ | $2.49 \times 10^{-7}$ | K.m |
| Diffusion coefficient of Mo in liquid Ni (constant) | $D^{(Mo)}$ | $1.60 \times 10^{-9}$ | m$^2$/s |
| Diffusion coefficient of Al in liquid Ni (Arrhenius) | $D_0^{(Ni)}$ | $1.86 \times 10^{-7}$ | m$^2$/s |
|  | $E_D^{(Ni)}$ | 63.68 | kJ/mol |
| Kinetic coefficient | $\mu$ | 0.7 | m/s/K |
| Solute drag coefficient | $\alpha$ | 0 or 24/35 | - |
| Interface width | $\delta$ | $10^{-9}$ | m |

## 4. Results

### *4.1. Direct Measurements of Solid-Liquid Interface Velocities*

The solidification velocities, shown in Fig. 3, tend to increase from the beginning to the end of solidification. Each plot is represented with a shaded background corresponding to a measurement error of ±1 pixel (± 1.93 μm) over the time between two successive frames (here with a frame rate of 67 200 or 80 000 Hz). Previous attempts at measuring growth velocities without applying the spatiotemporal smoothing procedure described in Section 3.1 led to noisy signals within these bounds. Lower laser powers tend to lead to higher solidification velocities. No significant difference in solidification velocities was observed as a function of crystallographic orientation.

The hatched region of Fig. 3 shows the velocity range above which the absolute stabilization of a planar interface is expected, according to classical theory (i.e., KGT, as detailed further below). Since solidification velocities measured for the lowest power (P = 104 W) clearly exceed this velocity range, one would expect the solid-liquid interface to re-stabilize, or at the very least the interfacial pattern to exhibit banded structures typical of the transition range between dendritic/cellular and planar at high growth velocities. For this reason, characterization results presented in Section 4.2 focus specifically on microstructures from the two experiments performed at the lowest laser power, i.e., those exhibiting the highest growth velocities (Fig. 3a).

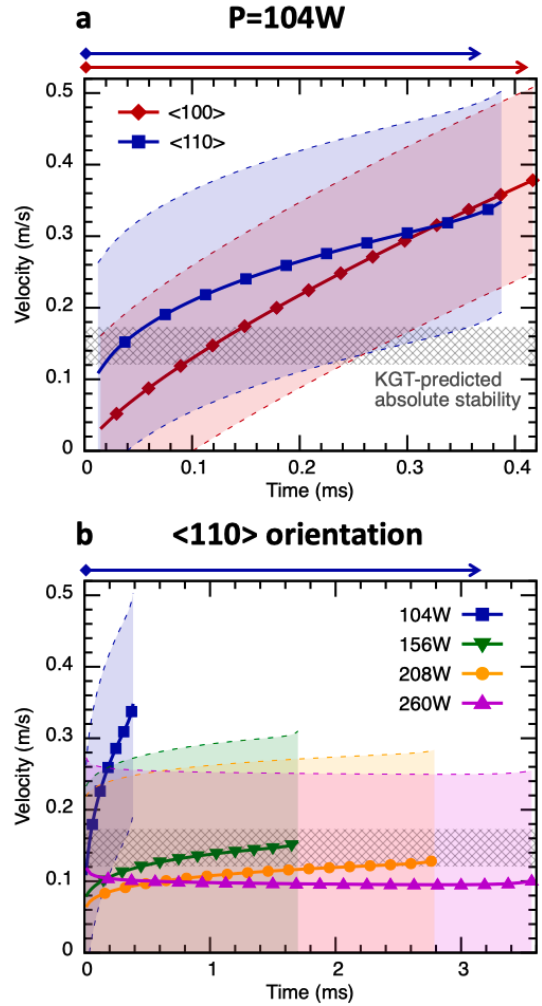

*Fig. 3: Solidification velocities measured from the bottom to the top of the melt pool (a) for a laser power of 104 W for different crystal orientations and (b) for one crystal orientation and different laser powers. The position from the edge of the melt pool is measured along the crystal ⟨100⟩ direction from the fusion line, as illustrated in Fig. 1c. The hatched region corresponds to the range of velocities at which absolute planar stability is expected to occur, according to the KGT model.*

### *4.2. Microstructures*

Microstructures observed by *ex-situ* characterization of the spot melts with P = 104 W are shown in Fig. 4. For both crystallographic orientations, they indicate an initial regime of planar growth at the bottom of the melt pool, i.e., at the onset of solidification, when growth velocities are low. As seen in Fig. 4b and 4e, the typical growth length of the initial planar interface is slightly below one micrometer. This is followed by a transition to cellular and columnar dendritic patterns, as growth velocities increase throughout the rest of the spot melt. As commonly observed in metals processed by AM, pronounced dendritic side-branches are absent, which is why they are often described as cellular. However, their growth kinetics and the mechanisms behind their morphological selection (e.g., tip radius) and preferred growth direction are indisputably that of dendrites. EBSD inverse pole figure (IPF) and image quality (IQ) maps in Fig. 4(a,d) indicate that the vast majority of the melt pool has the same crystallographic orientation as the substrate, and the main growth direction of cells or dendrites (annotated with black arrows in Fig. 2a and 2d) clearly correspond to ⟨100⟩ directions, as expected from dendrites, but not cells. Small areas near the top of the melt pools exhibit different crystallographic orientations, albeit still dendritic. These new orientations originate from nucleation ahead of the advancing solid-liquid interface, which may indicate the onset of an aborted columnar-to-equiaxed transition (CET) at the higher solidification velocities experienced at the top of the melt pool, although more work is needed to confirm this. What is decidedly absent in the microstructures of Fig. 4 is any indication of the re-stabilization of a planar interface, nor of any banding mechanism. Planar growth only occurs at the bottom of the melt pool (Fig. 4b and 4e), i.e., at the beginning of solidification, when the solid-liquid interface velocity is still below the limit of constitutional undercooling.

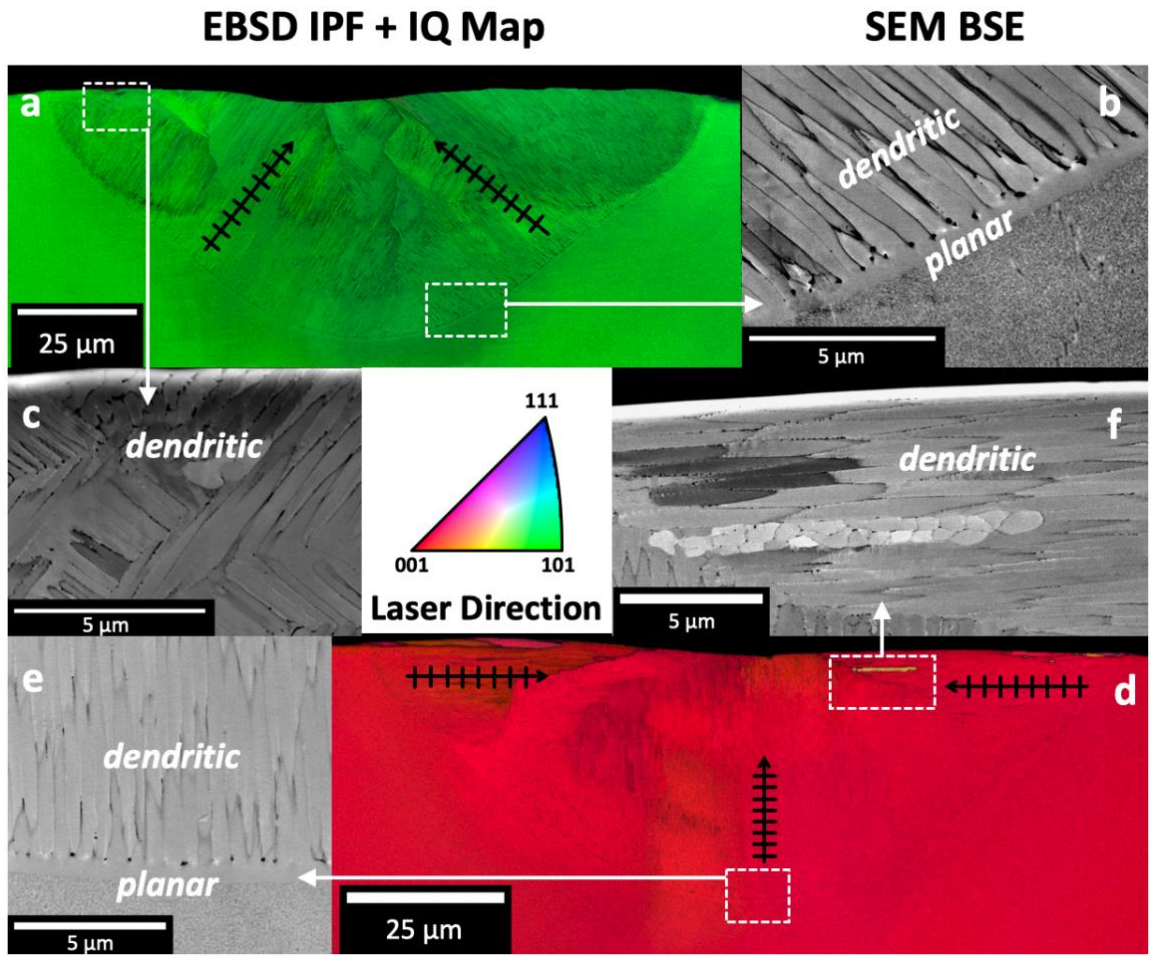

*Fig. 4: Ex-situ EBSD and SEM of cross-sections taken from ⟨100⟩ and ⟨110⟩ oriented samples (P = 104 W). (a) EBSD IPF and IQ map of a ⟨110⟩-oriented sample, with SEM backscattered electron (BSE) micrographs at the (b) bottom (c) and top of the top of the melt pool. (d) EBSD IPF + IQ map of a ⟨100⟩-oriented sample, with SEM BSE micrographs of the (e) bottom and (f) top of the melt pool.*

*4.3. Calculations*

The results of the KGT model (Eqs (5)-(9)) for the Ni-22.2Mo-2.8Al alloy at several temperature gradients from $10^4$ to $10^7$ K/m are compared to experimental results in Fig. 5. The temperature-velocity plot (Fig. 5a) also shows the velocity-dependent solidus and liquidus temperatures (Eqs (14)-(15)), considering no solute drag (solid lines) or complete solute drag (dashed lines). The local maxima of $T_S(V)$ at 0.8993 m/s (no drag) and 1.509 m/s (full drag) are marked with purple vertical lines below the curves. For the sake of clarity, KGT predictions for complete solute drag (black dashed line) and constant parameters estimated at $V = 0$ and $T = T_L$ (grey dotted line) are only plot for one temperature gradient ($G = 10^4$ K/m), because they only deviate from the no-drag curves at high velocities, at which G does not matter anymore and all curves collapse for each given set of assumptions (full drag, no drag, or constant parameters).

As expected, on the dendrite tip radius R(V) curves (Fig. 5b), the asymptotes with R→∞ are in good agreement with analytical estimations of both the constitutional undercooling limit, $V_c$, (Eq. (3)) and the absolute stability limit, $V_{abs}$, (Eq. (4)) marked with vertical dashed lines. Results are also shown using different assumptions, namely: (i) using constant parameters estimated at $T = T_L$ and $V = 0$, (ii) using temperature-dependent $D^{(i)}$ and velocity-dependent $k^{(i)}$ and $m^{(i)}$ without solute drag ($\alpha = 0$), or (iii) with complete solute drag ($\alpha = 24/35$). None of these assumptions dramatically change the onset of absolute stability $V_{abs}$. In fact, the absolute stability limit is even reduced from 0.173 m/s (constant parameters) to 0.120 m/s (no solute drag) or 0.122 m/s (full solute drag) when considering velocity-dependent parameters.

Superimposed in Fig. 5b are experimentally measured values of the primary dendrite spacing, $\lambda$ (Fig. 5d). The spacing is seen to follow the expected scaling $\lambda \sim R$ (with $\lambda/R$ independent of velocity and $\lambda$ a few times R) observed in other alloy systems [38,39]. The resulting stability plots in the G-V plane are compared in Fig. 5c to the maximum velocity measured in the laser melting/solidification experiments (dashed horizontal lines). These experimentally measured velocities appear above or close to the KGT predictions for $V_{abs}$, while they remain lower than those calculated using the $dT_S/dV = 0$ criterion to assess the absolute stability limit (upper shaded purple region).

## 5. Discussion

Measured solid-liquid interface velocities (Fig. 3) suggest that, according to KGT theory, absolute stability should have been reached towards the end of solidification in experiments performed at low laser power (P = 104 W). Yet, all microstructures observed by *ex-situ* microscopy (Fig. 4) exhibit clear solidification patterning, indicating that absolute stability was not achieved. Since *in-situ* radiographs at low laser power exhibit the lowest imaging contrasts, one could question the effect of the velocity extraction method (see Section 3.1 and Supplementary Material). However, the attached supplementary videos clearly show that the method appropriately captures the shape and evolution of the melt pools in each experiment. Moreover, a conservative estimation of the average velocity $\overline{V}$ of the solidification front, simply dividing the maximum radius of the melt pool by its total shrinking time lead to $\overline{V} \approx 0.241$ m/s for the ⟨100⟩-oriented sample (with r(90°) ≈ 108 μm solidified within 30 frames, i.e., 0.446 ms) and $\overline{V} \approx 0.252$ m/s for the ⟨110⟩-oriented sample (with r(45°) ≈ 104 μm within 33 frames, i.e., 0.4125 ms). Both of these values, which are much less prone to measurement artifacts, are significantly higher than the KGT-predicted absolute stability limit (at most 0.173 m/s for constant parameters). This unambiguously confirms that solidification rates in experiments performed at 104 W exceed the absolute stability limit estimated by the KGT model.

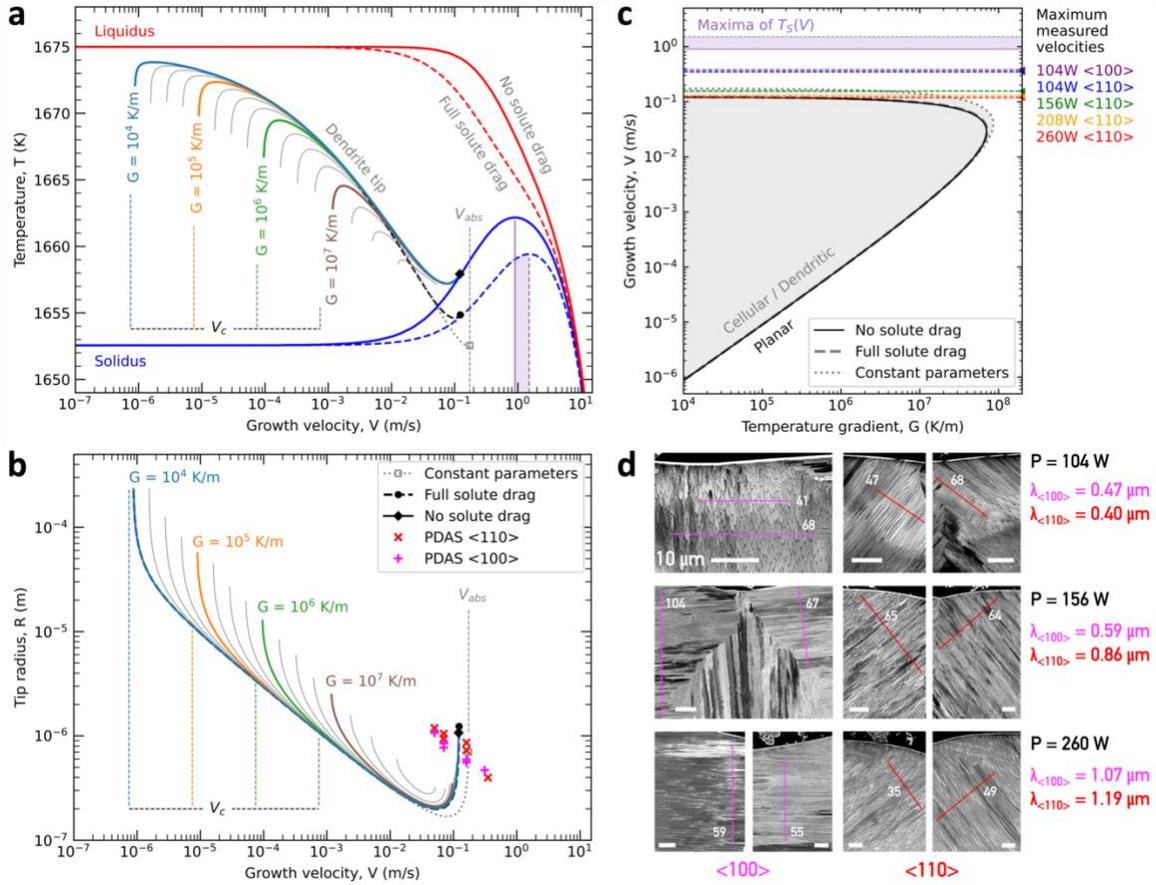

*Fig. 5: (a) Calculated tip temperature and (b) dendrite tip radius using the KGT model (Eqs (5)-(9)), considering different assumptions: (i) T- and V-dependent parameters without solute drag (solid lines; maximum V marked with a black diamond); (ii) T- and V-dependent parameters with complete solute drag (dashed line; maximum velocity: black circle); and (iii) constant parameters (dotted grey line; maximum velocity: open square symbol). Velocity-dependent solidus (blue) and liquidus (red) temperatures (Eqs (14)-(15)) also appear in (a), assuming no (solid line) or complete (dashed line) solute drag. Theoretical velocities for the limit of constitutional undercooling $V_c$ (Eq. (3)) and absolute planar stability $V_{abs}$ (Eq. (4)), are marked with vertical dashed lines. Panel (c) summarizes the resulting G-V stability map, compared to the maximum velocities measured in experiments (horizontal dashed lines). Panel (d) shows examples of microstructures used to measure the primary dendrite spacings (cross and plus symbols in (b)) for the two considered crystal orientations, with lines used to measure primary dendrite arm spacings (PDAS). Lines are labelled with the number of primary spacings counted within each line.*

The effect of the interface curvature can be assessed from the curvature undercooling $\Delta T_\kappa = \Gamma/R$. Using a dendrite tip radius of order 0.1 µm, as seen in the micrographs, and the Gibbs-Thomson coefficient listed in Table 1, the curvature undercooling is close to $\Delta T_\kappa = 2.49$ K. This is much lower than the variations of temperature linked to solute trapping or kinetic undercooling (Fig. 5a). This curvature undercooling can be observed (to a lower extent, because at lower V and hence higher R) by the distance between liquidus and tip temperature in Fig. 5a at low G and low V. The effect of the isotherm curvatures, with an initial radius of about 50 µm, is expected to have a negligible effect, as it is more than two orders of magnitude larger than the scale of the interface pattern, which makes the isotherms essentially (locally) planar from the standpoint of the interface pattern. As such, classical theories of pattern selection involving a one-dimensional temperature gradient (e.g., KGT) should still apply to these experiments. As a matter of fact, banding instabilities have been observed in several experiments involving curved isotherms, such as in welding [1,2] and *in-situ* dynamic TEM spot melting experiments [7,9].

While the temperature gradient could not be measured in these experiments, various simulations of laser powder bed fusion of Ni alloys show typical temperature gradients on the order of $10^7$ K/m or lower – e.g., $G \leq 1.84\times10^7$ K/m for P = 285 W [40], $G \leq 5\times10^7$ K/m for P = 162 to 201 W [41], or $G \leq 8\times10^6$ K/m for P = 100 W [42]. Preliminary spot-melting simulations performed for Ni alloys by the authors, which will be reported elsewhere, lead to $G \leq 2\times10^7$ K/m for $P \geq 100$ W, without considering fluid flow (alloy: Inconel-738, software *Sysweld*), and $G \leq 8\times10^6$ K/m for P = 104 W, accounting for fluid flow (alloy: Mo-lean ternary NiMoAl, software *Flow3D*). Such temperature gradients consistently identify the thermal conditions to be within the region where the limit of absolute stability is independent of G (Fig. 5c). This supports our conclusion that the considered G-V conditions should, according to classical KGT theory, lead to a restabilization of the planar interface, or at the very least a banding instability when approaching the absolute stability limit $V_{abs}$.

One limitation of the original KGT model is that it is oversimplified by assuming constant parameters. However, the use of temperature-dependent diffusivities and velocity-dependent partition coefficients and liquidus slopes does not substantially change the absolute stability threshold – it does, in fact, reduce it even further, consistent with original results in Ag-Cu alloys [19].

A second, more fundamental, limitation of the KGT model is that it is based on marginal stability theory, which assumes that the dendrite tip radius, R, is determined by the stability length, $\lambda_S$. The latter is defined as the length for which linear perturbations of the planar interface of wavelength larger (smaller) than $\lambda_S$ are unstable (stable). For this reason, $V_{abs}$ predicted by Eq. (4) coincides exactly with the velocity at which R diverges in Fig. 5(b), and beyond which the KGT theory predicts that dendrite growth is no longer possible. In contrast, microscopic solvability theory – the fundamentally correct self-consistent theory of dendrite tip operating state – predicts that R is determined by both the anisotropy of the excess free-energy of the interface [43], which plays a dominant role at slow solidification rates, and, in addition at rapid rates, the anisotropy of the interface kinetic coefficient µ [44,45]. Since neither of these anisotropies are included in the KGT model, it is theoretically possible for steady-state solutions of dendrite growth to persist beyond $V_{abs}$ predicted by Eq. (4), as observed here and in other alloy systems (e.g., Al-8wt%Fe, where rapid dendrite growth is observed for velocities much larger than $V_{abs}$ [38]).

In contrast, the estimation of the absolute stability velocity from the maximum of $T_S(V)$ leads to results that appear more consistent with the experimental observation of dendritic microstructures. Indeed, using this method, the lowest value of $V_{abs}$, namely 0.899 m/s (without solute drag), is more than twice the highest measured velocity in all experiments, namely 0.378 m/s for a laser power of 104 W in the ⟨100⟩ direction. Still, the observed microstructures do not show any indication of banding instability, and there is no known or theoretically obvious relationship between the maximum of the $T_S(V)$ curve and the largest allowed velocity for dendrite growth in the absence of banding. The absence of banding may indicate that noticeable solute trapping and absolute stability may be achieved even further than estimated here. The reduced sample size may also play a role, although banded microstructures have been shown in smaller TEM-sized samples at velocities approaching 2 m/s in binary Al alloys [7,9].

One limitation of the present modeling study is the consideration of separate independent solute contributions (no cross-coupling). In this respect, the selection of a model ternary alloy with one relatively dilute species (Al) should reduce the corresponding error. Moreover, although we have selected the alloy parameters with care, some discrepancies may still result from inaccurate thermophysical or kinetic parameters. Data for liquid state diffusivity is sparse for simple binary alloy systems, and largely nonexistent for more complex multicomponent alloy systems. Solid-liquid interfacial energies, needed to calculate the Gibbs-Thomson coefficient, as well as kinetic coefficient values are also essentially nonexistent from experimental measurements. Hence, we must rely on values calculated by atomistic (molecular dynamics) simulations, which, even though they are consistent with one another across various sources, remain arguably dependent on the choice of interatomic potentials and on their reliability near the melting temperature.

## 6. Conclusion

By performing laser spot melting experiments of single crystal Ni alloys with *in-situ* synchrotron X-ray imaging of melt pool solidification, we were able to accurately measure solidification velocities in AM-relevant conditions. Additional *ex-situ* characterization enabled unequivocal characterization of the solidified microstructures as branchless dendrites growing in preferred ⟨100⟩ crystal orientations, as expected for cubic crystal structures. Measured growth velocities are much higher than the absolute stability threshold predicted by classical theories, such as the KGT model, but none of the observed microstructures exhibit any sign of planar growth (except at low velocities in the early stages of melt pool shrinking), nor any sign of banding instability usually encountered when approaching the planar stability threshold in metal alloys. Moreover, modifications of the KGT model to account for the solid-liquid interface departure from equilibrium, namely using velocity-dependent partition coefficient and liquidus slope, reduce the expected threshold velocity for planar stability, hence further increasing the disagreement with experiments. We show that additional aspects, in particular the effect of kinetic undercooling, are required to predict an absolute stability threshold in agreement with experimental measurements. These results unambiguously show the shortcomings of classical – and widely used – theories in predicting solidification microstructures at high velocities, hence prompting for more accurate models to be employed in the AM community [46]. Finally, while we assessed alloy parameters as carefully as possible, it can be argued that further efforts are needed from the scientific community in the evaluation of accurate properties (e.g., liquid-state diffusivities, interface energies and kinetic coefficients and their anisotropies, etc.), without which sophisticated modeling frameworks will not have the fullest impact on our understanding of rapid solidification, even when coupled with state-of-the-art, *in-situ* diagnostics.

## Acknowledgements

Experiments were supported by the Department of the Navy, USA, Office of Naval Research, USA, under ONR award number N00014-18-1-2794. Any opinions, findings, and conclusions or recommendations expressed in this material are those of the author(s) and do not necessarily reflect the views of the Office of Naval Research. Modeling was supported by the U.S. Department of Energy, Office of Science, Office of Basic Energy Sciences under Award Number DE-SC0020870. This research used resources of the Advanced Photon Source, a U.S. Department of Energy (DOE) Office of Science User Facility operated for the DOE Office of Science by Argonne National Laboratory under Contract No. DE-AC02-06CH11357. DT gratefully acknowledges support from the Spanish Ministry of Science through a Ramón y Cajal Fellowship (Ref. RYC2019-028233-I).